\documentclass[preprint,tightenlines,aps,prd,groupedaddress,showpacs]
              {revtex4}%
\usepackage{graphicx}
\usepackage{amsmath}
\usepackage{bm}
\usepackage{relsize}
\RequirePackage{xspace}
\def\bea{\begin{eqnarray}}
\def\eea{\end{eqnarray}}
\def\nn{\nonumber}

\renewcommand\epsilon{\varepsilon}

\def\lsim{\mathrel{\raise.3ex\hbox{$<$\kern-.75em\lower1ex\hbox{$\sim$}}} }
\def\gsim{\mathrel{\raise.3ex\hbox{$>$\kern-.75em\lower1ex\hbox{$\sim$}}} }

\newcommand{\babar}{\mbox{\slshape B\kern-0.1em{\smaller A}\kern-0.1em %
B\kern-0.1em{\smaller A\kern-0.2em R}}\xspace}
\begin{document}

\title{\mbox{}\\[10pt]
Importance of Threshold Corrections \\
in Quark-Lepton Complementarity }

\author{Sin Kyu Kang}
\affiliation{ School of Physics, Seoul National University, Seoul
151-741, Korea}

\author{C. S. Kim and Jake Lee}
\affiliation{ Department of Physics, Yonsei University, Seoul 120-749,
Korea}



\date{\today}
\begin{abstract}
\noindent The recent experimental measurements of the solar neutrino mixing
angle $\theta_{sol}$ and the Cabibbo mixing angle $\theta_C$
reveal a surprising relation, $ \theta_{sol}+\theta_C \simeq
\frac{\pi}{4} $.
We note that the lepton mixing matrix derived from quark-lepton
unification can lead to a shift of the complementarity relation at
low energy.
While the renormalization group effects generally lead
to additive contribution on top of the shift, in this letter,
we show that the
threshold corrections which may exist in some intermediate scale
new physics such as supersymmetric standard model can diminish it,
so we can achieve the complementarity relation at a low energy.
Finally, we discuss a possibility to achieve the complementarity
relation at a high energy by taking particular form of
non-symmetric form of down Yukawa matrix.
\end{abstract}

%

\maketitle

Recently, it has been noted that the solar neutrino mixing angle
$\theta_{sol}$ required for a solution of the solar neutrino
problem and the Cabibbo angle $\theta_C$ reveal a surprising
relation \bea \theta_{sol}+\theta_C \simeq \frac{\pi}{4}, \eea
which is satisfied by the experimental results

$\theta_{sol}+\theta_{C}=45.4^{\circ}\pm 1.7^{\circ}$ to within a
few percent accuracy \cite{SK2002,SNO,fits}. This quark-lepton
complementarity (QLC) relation (1) has been interpreted  as an
evidence for certain quark-lepton symmetry or quark-lepton
unification as shown in Refs. \cite{raidal,smirnov,mohap}. Yet, it
can be a coincidence in the sense that reproducing the exact QLC
relation (1) at low energy scale  in the framework of grand
unification depends on the renormalization effects whose size can
vary with the choice of parameter space.  But, we believe that
such a coincidence does not necessarily mean that attempts to
catch a deep meaning behind the QLC relation were in vain.
%
Anyway, establishing the origin of the QLC relation may be a
challenge that underlying theory of flavor should address.

To effectively describe the deviation from maximal mixing of solar
neutrino, small mixing element $U_{e3}$ and possible deviation
from maximal mixing of atmospheric neutrino, a parametrization of
the PMNS mixing matrix in terms of a small parameter whose
magnitude can be interestingly around $\sin\theta_C$ has been
proposed as follows \cite{giunti, rodej,frampton,ramond}:
\bea
U_{\rm PMNS}=U^{\dagger}(\lambda)U_{\rm bimax}~. \label{framp}
\eea Here $U(\lambda)$ is a mixing matrix parameterized in terms
of a small parameter $\lambda$  and $U_{\rm bimax}$ corresponds to
the bi-maximal mixing matrix \cite{bimax}. Among possible generic
structures of the matrix $U(\lambda)$ which are compatible with
experimental results on the neutrino oscillations, the ``CKM-like"
form of $U(\lambda)$ has rather profound implication in view of
the connection between quarks and leptons.

In this letter, we first of all show that the lepton mixing matrix
given in the form of Eq. (\ref{framp}) with the ``CKM-like"
$U(\lambda)\sim U_{\rm CKM}$ can be indeed realized in the
framework of grand unification with symmetric Yukawa matrices when
we incorporate seesaw mechanism, and examine whether or not
$U_{\rm PMNS}$ reflecting quark-lepton unification given by
(\ref{framp}) can predict the QLC relation (1) exactly. We see
that the solar mixing angle derived from $U_{\rm PMNS}$ leads to
correction to the QLC relation which can be more than $1\sigma$
deviation from the QLC relation as similarly shown in
\cite{smirnov}. Notice that while the QLC relation holds at a low
energy, the corresponding relation derived from the mixing matrix
given in the form of Eq. (\ref{framp}) is in fact realized at a
high scale such as seesaw scale or unification scale. Thus, it is
necessary to take into account the renormalization effects on the
lepton mixing matrix when one compares the prediction at a high
energy scale with the QLC relation observed at low energy scale.
One can then expect that the deviation from the QLC relation is
explained by renormalization effects. In the SM, the
renormalization effect through the renormalization group (RG)
running down to the weak scale is negligible because of small
charged lepton Yukawa couplings. But, in some models such as the
minimal supersymmetric standard model (MSSM), the renormalization
effect on the leptonic mixing angle $\theta_{12}$ depends on the
type of light neutrino mass spectrum as well as model parameters
\cite{rgee0,rgee,rgee2}. It is also known that in MSSM with large
$\tan\beta $ and the quasi-degenerate neutrino mass spectrum the
RG effects are generally large and can enhance the mixing angle
$\theta_{12}$ at low energy \cite{rgee0,rgee,rgee2}. Such an
enhancement of $\theta_{12}$ is not suitable for achieving the QLC
relation (1) at low energy because we need to diminish the
deviation so as to get the exact QLC relation (1) at a low energy.

In this letter, we show that  the sizeable {\it threshold
corrections} which may exist in the MSSM
\cite{chan,slepton,neutuni} can diminish the deviation from the
QLC relation while keeping the mixing angle $\theta_{23}$ almost
maximal and $\theta_{13}$ small, so that the QLC relation at low
energy can be achieved in the case that the RG effects are
suppressed. Some conditions on the parameters to realize the QLC
relation will be discussed. Finally, we propose a possible way to
achieve the QLC relation by taking a non-symmetric form of the
down Yukawa matrix.

The flavor mixings stem from the mismatch between the left-handed
rotations of the up-type and down-type quarks, and the charged
leptons and neutrinos.
For our purpose, it is
useful to work in a basis where the quark and lepton Yukawa
matrices are related. In general, the quark Yukawa matrices
$Y_u,Y_d$ are given by
$Y_u=U_uY_u^{diag}V_u^{\dagger},~~Y_d=U_dY_d^{diag}V_d^{\dagger}$,
from which the observable quark mixing matrix is deduced as
$U_{\rm CKM}=U^{\dagger}_uU_d~$. 
For the charged
lepton sector, the Yukawa matrix is given by
$Y_l = U_lY_l^{diag}V^{\dagger}_l~$.
For the neutrino sector, we introduce one right-handed singlet
neutrino per family which leads to the seesaw mechanism, according
to which the light neutrino mass matrix after the breakdown of the
electroweak symmetry is given by
\bea
M_{\nu} &=& M_{\rm Dirac}\frac{1}{M_R}M_{\rm Dirac}^T \nn \\
&=& \left( U_{0}M_{\rm Dirac}^{diag}V^{\dagger}_0 \right)
\frac{1}{M_R} \left(V^{\ast}_0 M_{\rm Dirac}^{diag} U^T_0
\right),\label{lep1} \eea where $U_0$ and $V_0$ are the
left-handed and right-handed mixing matrices of the Dirac neutrino
mass matrix, respectively. We can then rewrite $M_{\nu}$ as
follows
 \bea M_{\nu} &=& U_0V_M M_{\nu}^{diag}V_M^T U_0^T,
\label{lep2} \eea
where $V_M$ represents the rotation of $M_{\rm
Dirac}^{diag}V^{\dagger}_0 \frac{1}{M_R}V^{\ast}_0 M_{\rm
Dirac}^{diag}$~.
 The observable PMNS mixing matrix can then be
written as \cite{ramond} \bea U_{\rm PMNS} = U^{\dagger}_l U_{\nu}
= U^{\dagger}_lU_0V_M~.\label{lep3} \eea

Note that equating the above expression
for $U_{\rm PMNS}$ with  Eq. (\ref{framp}),
we get the
``CKM-like" form of $U(\lambda)$:
\bea
 U^{\dagger}(\lambda)=U^{\dagger}_lU_0V_MU^{\dagger}_{\rm bimax}.
 \label{eq}
\eea
In order for $U^{\dagger}(\lambda)$ to have ``CKM-like" small mixing angles
we have three
 possible choices of $U^{\dagger}(\lambda)$:
 \bea
U^{\dagger}(\lambda)=\left\{\begin{array}{l} U_l^{\dagger}~, \\
U^{\dagger}_lU_0~ , \\
U^{\dagger}_lU_0V^{\prime}~, \end{array}\right. \eea
where
$V^{\prime}=V_MU^{\dagger}_{\rm bimax}$. The first choice of
$U^{\dagger}(\lambda)$ indicates that the maximal mixing angles in
$U_{\rm bimax}$ are cancelled out by the mixing angles in the
combination of $U^0 V_M$, while the second choice implies $V_M=U_{\rm
bimax}$. The last form is the most general one, which shows that the
maximal mixing angles in $U_{\rm bimax}$ are not completely
cancelled out, and its actual form is not unique.
In view of the quark-lepton unification, the second case is more
natural than others because down-type (up-type) quarks are related
with charged lepton (Dirac neutrino) sector in grand unification.
In such a case, the form of $V_M$ is taken to be almost bi-maximal
mixing because it is natural to suppose that
the structure of the lepton mixing matrix to a leading order is the bi-maximal mixing,
whereas the CKM matrix is an identity matrix, which corresponds
to $U^{\dagger}(\lambda)= {\rm I}$ in Eq. (\ref{framp}), and then
the QLC relation can emerge from quark-lepton unification.
However, the bi-maximal mixing pattern of
$V_M$ is not necessarily required.
It is in fact possible to take $V_M$ to be small mixing or even identity
matrix at GUT scale and then to generate two large mixing angles
in the lepton mixing matrix by radiative magnification through
evolving RG equations down to the weak scale \cite{rgee,rgee2}.

Now, let us consider how
the PMNS mixing matrix given by Eq. (\ref{lep3}) can be related
with CKM mixing matrix in the context of quark-lepton unification.


{\bf (A) Minimal quark-lepton unification}\\
Since the down-type quarks and the charged leptons are in general
assigned into a multiplet in grand unification, we assume that the
following simple relations hold in the minimal models of grand
unification,
$Y_e=Y^T_d,~~~Y_u=Y^T_u$.
Then, we deduce that
$U_l=V_d^{\ast}$ from which \bea
  U_{\rm PMNS}=V^T_d U_0 V_M.
\eea
{}From this expression for $U_{\rm PMNS}$, we see that
the contribution of $U_{\rm CKM}$ may appear in $U_{\rm PMNS}$ if
we further assume $Y_{\nu}=Y_u$ which can be realized in some
larger unified gauge group such as $SO(10)$. Then, one can obtain
\cite{ramond} \bea U_{\rm PMNS}
= V^T_d U_d U_{\rm CKM}^{\dagger}V_M. \eea In addition, requiring
symmetric form of the down-type quark Yukawa matrix, we finally
obtain \bea U_{\rm PMNS}=U^{\dagger}_{\rm CKM}V_M, \label{sym}\
\eea where the mixing matrix $V_M$ has bi-maximal mixing pattern
as explained above. In this way, $U_{\rm PMNS}$ can be connected
with $U_{\rm CKM}$. We note that taking the bi-maximal mixing form
of $V_M$ is equivalent to taking $M_R$ as follows, in the basis
where $V_0$ is absorbed into the heavy Majorana neutrino field:
\bea M_R=\frac{1}{2}\left(\begin{array}{ccc} \alpha m_{D_1}^2 & \beta
m_{D_1}m_{D_2} & -\beta m_{D_1}m_{D_3} \\
\beta m_{D_1}m_{D_2} & \left(\frac{1}{2}\alpha + \gamma
\right)m^2_{D_2} & \left(-\frac{1}{2}\alpha + \gamma
\right)m_{D_2}m_{D_3} \\
-\beta m_{D_1}m_{D_3} & \left(-\frac{1}{2}\alpha + \gamma
\right)m_{D_2}m_{D_3} & \left(\frac{1}{2}\alpha + \gamma
\right)m^2_{D_3} \end{array}\right)~,
\eea
where
$\alpha=m_1^{-1}+m_2^{-1}, \beta=(-m_1^{-1}+m_2^{-1})/\sqrt{2},
\gamma=m_3^{-1}$ and $m_{D_i}, m_i$ stand for the mass eigenvalues
of Dirac mass matrix and light neutrino mass matrix, respectively.

To see whether the parametrization of $U_{\rm PMNS}$ given by
(\ref{sym}) can lead to the QLC relation (1), it is convenient to
present $U_{\rm PMNS}$ for the CP-conserving case as follows:
\bea
U_{\rm PMNS} &=& U^{\dagger}_{\rm CKM}U_{23}^m U_{12}^m
\nonumber \\
&\equiv&
U_{23}(\theta_{23})U_{13}(\theta_{13})U_{12}(\frac{\pi}{4}-\theta_{12})~,
\label{ckm3} \eea where $U_{12}^m$ and $U_{23}^m$ correspond to
the maximal mixing between (1,2) and (2,3) generations,
respectively. Then, the mixing angles $\theta_{ij}$ in the second
line of Eq. (\ref{ckm3}) can be presented in terms of Wolfenstein
parameter $\lambda$ as follows: \bea \sin\theta_{12} &\simeq &
\frac{1}{\sqrt{2}}\lambda+O(\lambda^3), ~~~ \sin\theta_{23} \simeq
-\frac{1}{\sqrt{2}}\left(1-\frac{1}{2}\lambda^2\right), ~~~
\nonumber \\
\sin\theta_{13} &\simeq &  -\frac{1}{\sqrt{2}}\lambda~. \eea The
solar neutrino mixing parameter $\sin\theta_{sol}$ in this
parametrization becomes \bea \sin\theta_{sol}\simeq
\sin\left(\frac{\pi}{4}-\theta_C\right)
+\frac{\lambda}{2}(\sqrt{2}-1). \label{pred} \eea Thus, we see
that the neutrino mixing matrix (\ref{ckm3}) originating from the
quark-lepton unification obviously leads to a shift of the
relation (1). Numerically, the shift amounts to $\delta
\theta_{sol} \simeq 3^{\circ}$ and thus $\delta
\sin^2\theta_{sol}\simeq 0.05$ which is more than $1\sigma$
deviation from the recent measurement of the solar neutrino
experiment. While a dedicated experiment to measure $\theta_{12}$
with a sensitivity of a few $\%$ to $\sin^2\theta_{12}$ would be
expected to confirm or rule out the deviation, we can expect that
renormalization effects on the neutrino mixing matrix (\ref{ckm3})
may fill the gap between the QLC relation and the prediction
(\ref{pred})from high energy mixing matrix. Later, we will discuss
the renormalization effect in detail.

{\bf (B) Realistic quark-lepton unification } \\
Although the minimal quark-lepton unification can lead to an
elegant relation between PMNS mixing  and CKM mixing as shown in
the above, it indicates undesirable mass relations between quarks
and leptons at the GUT scale such as $m_d^{diag}=m_l^{diag}$.
Recently, the following form for
$U^{\dagger}_lU_0=U^{\dagger}(\lambda)$ has been suggested based
on a well known empirical relation $|V_{us}| \simeq
\sqrt{\frac{m_d}{m_s}}\simeq 3\sqrt{\frac{m_e}{m_{\mu}}}$
\cite{pakvasa}, \bea U^{\dagger}(\lambda)=U^{\dagger}_lU_0\simeq
\left(\begin{array}{ccc}
 1 & -\frac{\lambda}{3} & \frac{1}{3}\theta\lambda^2 \\
\frac{\lambda}{3} & 1 &
2\theta\lambda \\
-\theta \lambda^2 & -2\theta \lambda &
1\end{array}\right)\label{real}, \eea where the deviation from
unitarity is just of order $\theta \lambda^3$. This form of mixing
matrix can be obtained by introducing the Higgs sector
transforming under the representation {\bf 45} of $SU(5)$ or {\bf
126} of $SO(10)$ \cite{higgs}. In this case,
the solar neutrino mixing $\sin\theta_{sol}$ is then given by \bea
\sin\theta_{sol} \simeq \sin\left(\frac{\pi}{4}-\theta_C\right)
+\frac{\lambda}{2}\left(\sqrt{2}-\frac{1}{3}\right)~. \eea
Numerically, the deviation from the QLC relation amounts to
$\delta \theta_{sol}\simeq 7^{\circ}$, much more than in the
observed QLC relation. One possible way to generate the correct
prediction for $\sin\theta_{sol}$ based on the lepton mixing with
$U_l$ given by Eq. (\ref{real}) is to abandon the exact bimaximal
form of the neutrino mixing matrix $V_M$ and to consider the
generic corrections to the bimaximal neutrino mixing matrix that
can account for the QLC relation \cite{pakvasa}.
But, we consider an alternative possibility that the threshold
corrections can diminish the deviation from the QLC relation.

 Now, let us examine how the renormalization effects can diminish
 the deviation from the QLC relation.  In general, the radiative
corrections to the effective neutrino mass matrix can be presented
as follows:
\bea
M_{\nu} &=& I\cdot M^0_{\nu} \cdot I \nonumber \\
&=& I\cdot U^T_{\rm CKM}U^{\ast}_{\rm bimax}M_DU^{\dagger}_{\rm bimax}
U_{\rm CKM}\cdot I\nonumber\\
&=& I\cdot U^T_{\rm CKM}U_{23}^{m^{\ast}} M_{D12}U^{m\dagger}_{
23}U_{\rm CKM} \cdot I~,
\eea
where  $M_D=Diag[m_1,m_2,m_3]$, $M_{D
12}=U^{m^{\ast}}_{12} M_D U^{m\dagger}_{12}$, and the matrix
$I\equiv I_A\delta_{AB}, (A,B = e, \mu, \tau)$ stands for the
radiative corrections. The correction $I$ generally consists of
two parts $I=I^{RG}+I^{TH}$ where $I^{RG}$ are the renormalization
group corrections which are explicitly presented in Ref.
\cite{ellis} and $I^{TH}$ are electroweak scale threshold
corrections \cite{chan}. We note that the flavor blind
interactions such as gauge interactions contribute to overall
scale of neutrino masses whereas the charged lepton Yukawa
interactions generate flavor dependent radiative corrections $I$
in the standard model (SM) and in MSSM. The typical size of RG
corrections $I^{RG}$ is known to be about $10^{-6}$ in the SM and
MSSM with small $\tan\beta$, and thus negligible. In addition,
supersymmetry can induce flavor dependent threshold corrections
related with slepton mass splitting which can dominate over the
charged lepton Yukawa corrections \cite{slepton}. Even if
$I_{\tau}$ is the dominant contribution in SM, it is not
guaranteed at all in MSSM due to the threshold corrections. We
have numerically checked that RG evolution from the seesaw scale
to the weak scale {\it enhances} the size of $\theta_{12}$ in the
case that $\theta_{13}$ and $\theta_{23}$ are kept to be small and
almost maximal mixing, respectively. Thus, the case of sizable RG
effects is not suitable for our purpose. Instead, we examine
whether the threshold corrections can be suitable for diminishing
the deviation from the QLC relation while keeping $\theta_{23}$
nearly maximal and $\theta_{13}$ small in the case that RG effect
is negligible.
 To achieve our goal, we note that the contribution
$I_e$ should be dominant over $I_{\mu,\tau}$ because only $I_e$
can lead to the right amount of the shift of $\theta_{12}$ while
keeping the changes of $\theta_{23}$ and $\theta_{13}$ small.
Taking $|I_e| >> |I_{\mu,\tau}|$, the neutrino mass matrix
corrected by the leading contributions is  rewritten as follows:
 \bea M_{\nu}
 &\simeq& U^{T}_{\rm CKM}U_{23}^{m^{\ast}}\left[I_D+I_e
 \Lambda_{\lambda}^{\ast}
 \right]M_{D 12}\left[I_D+I_e\Lambda^{\dagger}_{\lambda}\right]
 U^{m\dagger}_{23} U_{\rm CKM}~, \label{corr}
 \eea
 where $I_D$ is $3\times 3$ identity matrix, and
 the matrix $\Lambda_{\lambda}$ is given up to $\lambda^2$ order by
\bea
\Lambda_{\lambda} &=& \left(\begin{array}{ccc}
 1 & -\frac{\lambda}{\sqrt{2}} & -\frac{\lambda}{\sqrt{2}} \\
-\frac{\lambda}{\sqrt{2}} & \frac{\lambda^2}{2} &
\frac{\lambda^2}{2} \\
-\frac{\lambda}{\sqrt{2}} & \frac{\lambda^2}{2} &
\frac{\lambda^2}{2}\end{array}\right)  ~~\mbox{for minimal
unification,} \\
&=& \left(\begin{array}{ccc}
 1 & -\frac{\lambda}{3\sqrt{2}} & -\frac{\lambda}{3\sqrt{2}} \\
-\frac{\lambda}{3\sqrt{2}} & \frac{\lambda^2}{18} &
\frac{\lambda^2}{18} \\
-\frac{\lambda}{3\sqrt{2}} & \frac{\lambda^2}{18} &
\frac{\lambda^2}{18}\end{array}\right)  ~~\mbox{for realistic
unification.} \eea

Let us discuss how much the lepton mixing angles can be shifted by
the renormalization effects. We first of all do numerical analysis
in a model independent way based on the form given by Eqs. (19--20).
For our numerical analysis, we take $\Delta
m^2_{sol}\equiv m_2^2-m_1^2\simeq 7.1\times 10^{-5}~\mbox{eV}^2$
and $\Delta m^2_{atm}\equiv m^2_3-m_2^2\simeq 2.5\times
10^{-3}~\mbox{eV}^2$. Regarding $M_{\nu}$ as the light neutrino
mass matrix at low energy scale and varying the parameter $I_e$
and the smallest light neutrino mass $m_1$, we find which
parameter set $(I_e, m_1)$ can lead to the QLC relation exactly
and the results are presented in Table I. We note that the
required values for $I_e$ are negative. The first and the second
column in Table I correspond to the minimal unification and
realistic case, respectively.
 In our analysis, we have also checked that
 $\theta_{23}$ is almost unchanged,
whereas  the shift of $\theta_{13}$ is about $1^\circ$ for both
cases (A) and (B). {}From Table I, we see that a larger value of
$I_e$ is required to achieve the relation (1) as $m_1$ goes down.
\begin{table}
\begin{tabular}{|c|c|c|}\hline
 & (A) & (B)  \\
\hline
 $m_1$ (eV) & $I_e$ & $I_e$ \\
 \hline \hline
 0.15 & $-4.0\times 10^{-5}$ &  $-1.0\times 10^{-4}$\\
 0.1 & $-8.5\times 10^{-5}$ &  $-2.2\times 10^{-4}$\\
 0.05 &  $-3.4\times 10^{-4}$&  $-8.6\times 10^{-4}$\\
\hline
\end{tabular}
 \caption
 {
 Parameter set $(I_e, m_1)$ leading to the QLC relation.
 The columns (A) and (B) correspond to the minimal unification and
 realistic case, respectively.
 }
\end{table}

How can we obtain such a value of $I_e$ while keeping
$|I_e|>>|I_{\mu,\tau}|$? As shown in Ref. \cite{neutuni}, it
requires the existence of new states which gives a dominant
contributions to $I_e$. In MSSM, it can easily be realized by
taking into account the contribution of chargino (pure W-ino). In
the case of a diagonal slepton mass matrix in the same basis where
the charged lepton mass matrix is diagonal, the contribution is
presented by \cite{neutuni} \bea I_{f} \simeq
\frac{g^2}{32\pi^2}\left(-\frac{1}{x_f}+\frac{1}{x_{\mu}}+\frac{x_f^2-1}{x_f^2}\ln(1-x_f)
-\frac{x_{\mu}^2-1}{x_{\mu}^2}\ln(1-x_{\mu})\right)~, \eea where
$x_{f}=1-(M_{f}/\tilde{m})^2$ and $M_{f}, \tilde{m}$ are the
$f$-th charged slepton mass and W-ino mass, respectively. We see
that the size of $|I_e|$ is about 10 times larger than that of
$|I_{\mu,\tau}|$ for
 $M_e\sim 2M_{\mu,\tau}$, and the value of $I_e$ becomes negative and of the order
 of $10^{-4} \sim 10^{-3}$ for
 $x_e\gtrsim 0.65$, which are required to achieve the exact QLC relation at low energy.
 We also note that
the contribution of the tau Yukawa coupling $Y_{\tau}$ in MSSM to
$I_{\tau}$ can be as much as $I_e$ for large $\tan\beta$. Thus,
our estimation is suitable for small $\tan \beta$ so that the
contribution of $Y_{\tau}$ should be negligible compared to $I_e$.

In passing, we discuss a possible way to diminish the correction
to QLC appeared in Eq. (\ref{pred}). While we have considered the
case with symmetric form of fermion Yukawa matrices so far,
non-symmetric types of $Y_d$ and $Y_l$ are generally allowed.
Then, the term $V_d^TU_d$ in Eq. (9) is not trivial unity matrix
in the non-symmetric Yukawa structure and it can generate
additional correction to the solar neutrino mixing parameter
$\sin\theta_{sol}$, which can diminish the correction to QLC.  We
have checked that $\sin\theta_{sol}\simeq \sin(\pi/4-\theta_C)$
for the case of the minimal unification (A) when $V_d^TU_d$ has
the form in the leading order of $\lambda$, \bea V_d^TU_d=
\left(\begin{array}{ccc}
 1 & (1-\sqrt{2})\lambda & -(1-\sqrt{2})\lambda \\
-(1-\sqrt{2})\lambda & 1 & 0 \\
(1-\sqrt{2})\lambda & 0 & 1 \end{array}\right). \label{corr}
\eea
For the realistic case (B), one can obtain the QLC relation by
taking the matrix form (\ref{corr}) but replacing $\lambda$ with
$\lambda/3$. We note that in order to achieve the QLC relation at
low energy scale in those cases, renormalization effects should be
negligibly small. But, it is quite arbitrary to generate the form
of $V_d^TU_d$ given above from some underlying symmetries
or flavor structure.

In summary, while the QLC relation, $
\theta_{sol}+\theta_C=\frac{\pi}{4} $, itself can be an evidence
for the quark-lepton unification, it can be a coincidence in the
sense that the relation achieved at low energy in the framework of
grand unification strongly depends on the renormalization effects
whose size can vary with the choice of parameter space. In this
paper, we have found that the lepton mixing matrix derived from
quark-lepton unification can lead to a shift of the
complementarity relation at a high energy. While the
renormalization group effects generally lead to additive
contribution on top of the shift, we show that the threshold
corrections which may exist in the supersymmetric standard model
diminish it, so we can achieve the complementarity relation at a
low energy. In addition, we commented on how to achieve the QLC
relation by taking a non-symmetric form of the down Yukawa matrix.
\\

We would like to thank G. Cvetic for careful reading of the manuscript
and his valuable comments.
The work of S.K.K was supported in part by BK21 program of the Ministry
of Education in Korea and in part by  KOSEF grant
R02-2003-000-10085-0. The work of C.S.K. was supported in part by
CHEP-SRC Program and in part by Grant No. R02-2003-000-10050-0 from
BRP of the KOSEF.  The work of
J.L. was supported in part by BK21 program of the Ministry of Education
in Korea and in part by Grant No. F01-2004-000-10292-0 of
KOSEF-NSFC International Collaborative Research Grant.


\end{document}